\documentclass[10pt,conference]{IEEEtran}
\usepackage[dvips]{graphicx}
\usepackage{amsmath,amssymb,color,verbatim}

\def\norm#1{\left\| #1 \right\|}
\newtheorem{definition}{Definition}[section]
\newtheorem{thm}{Theorem}[section]
\newtheorem{proposition}[thm]{Proposition}
\newtheorem{lemma}[thm]{Lemma}
\newtheorem{corollary}[thm]{Corollary}
\newtheorem{exam}{Example}[section]

\renewcommand{\_}{\underline}

\def\bea{\begin{IEEEeqnarray}{rCl}} 
\def\eea{\end{IEEEeqnarray}}
\def\beq{\begin{equation}}
\def\eeq{\end{equation}}
\def\bean{\begin{IEEEeqnarray*}{rCl}} 
\def\eean{\end{IEEEeqnarray*}}

\newtheorem{remark}{Remark}[section]

\DeclareMathOperator*{\tr}{Tr}

\providecommand{\abs}[1]{\ensuremath{\left\lvert #1 \right\rvert}}
\providecommand{\norm}[1]{\ensuremath{\left\Vert #1 \right\Vert}}

\providecommand{\ceil}[1]{\ensuremath{\left\lceil #1 \right \rceil}}

\providecommand{\vv}[1]{\textquotedblleft #1\textquotedblright}

\newcommand{\Z}{\mathbb{Z}}
\newcommand{\C}{\mathbb{C}}
\newcommand{\R}{\mathbb{R}}

\renewcommand{\IEEEQED}{\IEEEQEDopen}

\providecommand{\abs}[1]{\ensuremath{\left\lvert #1 \right\rvert}}
\providecommand{\norm}[1]{\ensuremath{\left\Vert #1 \right\Vert}}

\providecommand{\vv}[1]{\textquotedblleft #1\textquotedblright}

\begin{document}

%\title{ Shifted Inverse Determinant Sum is a versatile tool for analyzing codes in fading channels}
\title{Shifted inverse determinant sums and new bounds for the DMT of space-time lattice codes}
\author{
\IEEEauthorblockN{Roope Vehkalahti}
\IEEEauthorblockA{Department of  mathematics and statistics\\
University of Turku\\
Finland\\
 roiive@utu.fi}
\and
\IEEEauthorblockN{Laura Luzzi}
\IEEEauthorblockA{Laboratoire ETIS\\
 CNRS - ENSEA - UCP \\
Cergy-Pontoise, France \\
laura.luzzi@ensea.fr}
\and
\IEEEauthorblockN{ Jean-Claude Belfiore}
\IEEEauthorblockA{Dept of Communications and Electronics\\
 Telecom ParisTech \\
Paris, France \\
belfiore@enst.fr}

}

\maketitle

\begin{abstract}
This paper considers shifted inverse determinant sums arising from the
union bound of the pairwise error probability for space-time codes in
multiple-antenna fading channels.
Previous work by Vehkalahti et al. focused on the approximation of these
sums for low multiplexing gains, providing a complete classification of
the inverse determinant sums as a function of constellation size for the
most well-known algebraic space-time codes.
This work aims at building a general framework for the study of the
shifted sums for all multiplexing gains. New bounds obtained using dyadic
summing techniques suggest that the behavior of the shifted sums does
characterize many properties of a lattice code such as the
diversity-multiplexing gain trade-off, both under maximum-likelihood
decoding and infinite lattice naive decoding. Moreover, these bounds allow
to characterize the signal-to-noise ratio thresholds corresponding to
different diversity gains.

%When considering lattice codes for the Gaussian channel, combining the union bound with an expression for the pair-wise error probability allows  one to upperbound the error probability by an exponential sum (theta series) over the lattice points. There is an extensive literature concerning this type of sums, both in mathematics and engineering.\\
%In fading multiple-antenna channels the role of the 
%theta series
%%exponential sum 
%is played  by  shifted inverse determinant sums. However, their behavior is not so well understood in the literature.
%%previous literature on the topic of these sums. 
%
%Unlike  their high signal-to noise approximations studied earlier by Vehkalahti \emph{et al.}, shifted determinant sums can be used to analyze any kind of space-time lattice code, not just the ones having the non-vanishing determinant property.\\
%This paper concentrates on building a general framework for the study of these sums. Results given suggest that their behavior 
%%of the shifted inverse determinant sum 
%does characterize many properties of a lattice code such as the diversity-multiplexing gain trade-off.

\end{abstract}

\section{Introduction}
Shifted inverse determinant sums appear naturally when analyzing the union bound for the pairwise error probability (PEP) of space-time codes  over MIMO channels \cite{TSC}. 
The high-SNR approximation of these sums was analyzed in \cite{VLL2013}, providing general bounds on the performance of algebraic space-time codes from division algebras and number fields. In particular, it was shown that the approximate sums are enough to characterize the diversity-multiplexing gain trade-off (DMT) \cite{ZT} of these codes in the multiplexing gain range $r \in [0,1]$. However, in order to study the DMT for higher multiplexing gains $r$, it becomes necessary to consider the original shifted determinant sums. 
%This need  came is evident in the work of Belfiore and Oggier in \cite{BO}, where the authors analyzed Alamouti type codes in  a wiretap MIMO channel.
In this work we provide a general framework to analyze shifted sums, which are able to predict the correct DMT curve for $r>1$ in some cases. We also discuss the characterization of the \vv{high SNR} threshold as a function of constellation size. \\ 
Moreover, we show that while their high-SNR approximations never converge, shifted sums always converge if the number of receive antennas is large enough; this provides new bounds on the DMT performance of naive lattice decoding.

Inverse determinant sums in the sense we are discussing were considered by Tavildar and Viswanath in \cite{TV}, where the authors analyzed the DMT of several simple space-time codes. 
The  most recent appearance of these sums is in the work of Belfiore and Oggier concerning
%Another problem in which shifted determinant sums play a role is the analysis of 
the eavesdropper's error probability in the MIMO wiretap channel \cite{BO}.\\
%, where the importance of considering the whole shifted sums was also stressed. 
Our take on the subject follows the general setting of \cite{VLL2013}, but we replace the approximation of PEP by the more accurate version. The idea to  consider symmetric polynomials and their relations to analyze PEP was given in  \cite{VH}.

\subsection{Matrix Lattices and spherically shaped coding schemes}\label{latticesection}
Before we can introduce inverse determinant sums, we need a few definitions.

\begin{definition}
A {\em matrix lattice} $L \subseteq M_{n\times T}(\C)$ has the form
$$
L=\Z B_1\oplus \Z B_2\oplus \cdots \oplus \Z B_k,
$$
where the matrices $B_1,\dots, B_k$ are linearly independent over $\R$, i.e., form a lattice basis, and $k$ is
called the \emph{rank}  or the \emph{dimension} of the lattice.
\end{definition}

\begin{definition}\label{def:NVD}
If the minimum determinant of the lattice $L \subseteq M_{n \times T}(\C)$ is non-zero, i.e. 
$\inf_{{\bf 0} \neq X \in L} \abs{\det (XX^*)} > 0, $
we say that the lattice satisfies the \emph{non-vanishing determinant} (NVD) property.
\end{definition}

We now consider a spherical shaping scheme based on a $k$-dimensional lattice $L$ in $M_{n\times T}(\C)$. 
Given $M>0$ we define
$$
L(M)=\{a \in L \;:\; \norm{a}_F \leq M, a\neq {\bf 0} \}.
$$
Here $\norm{\cdot}_F$ refers to the Frobenius norm.

\subsection{Motivation and problem statement}
Let us suppose that we are considering the complex  Gaussian channel and a finite code $L(M)\in \C^n$. 
If the codewords are sent equiprobably, we can upper bound the average error probability 
%(slightly simplifying)  
by the  sum
\[
P_e \leq \sum_{\_{x} \in L,\, 0 < \norm{\_{x}}_E \leq 2M} e^{-\norm{\_{x}}^2},
\]
where the  term $2M$ follows from the fact that we have to consider differences of codewords. The right-hand-side is then a well known truncated \emph{exponential sum} taking values on lattice points of $L$. Let us now describe the analogous bound in the fading channel.\\
Suppose that we have a lattice $L\subset M_{n\times T}(\C)$ and that we have chosen a finite code $L(M)$ and a constant $\theta$ such that $\theta L(M)$ has average energy $1$. \\
Consider the Rayleigh block fading MIMO channel with $n=n_t$ transmit and $n_r$ receive antennas. The channel is assumed to be fixed for a block of $T$ channel uses, but to vary in an independent and identically distributed (i.i.d.) fashion from one block to another. Thus, the channel input-output relation can be written as 
\beq
Y=\sqrt{\frac{\rho}{n}}H\theta X +N, \label{eq:channel}
\eeq
where $H \in M_{n_r \times n}(\C)$ is the channel matrix and $N\in M_{n_r \times T}(\C) $ is the noise matrix. The entries of $H$ and $N$ are assumed to be  i.i.d. zero-mean complex circular symmetric Gaussian random variables with variance 1.  
The matrix $X \in L(M)$ is the transmitted codeword, and the term $\rho$ denotes the signal-to-noise ratio (SNR). 

Following \cite{TSC},  we can upper bound the pairwise error probability between two codewords $X \neq X'$, when transmitting with SNR $\rho$, as follows:
\[
P(\rho,X \to X')\leq \frac{1}{(\det(I+ \frac{\rho\theta^2}{4n}(X-X')(X-X')^*))^{n_r}}, 
\]
where $*$ denotes complex conjugate transpose. The scaling factor $4n$ for the SNR $\rho$ is irrelevant for our asymptotic analysis so we will omit it in the sequel. \\ 
We can upperbound the average error probability, when transmitting a codeword from $L(M)$, as
$$
P_e\leq  \sum_{X\in L,\; 0 <||X||_F \leq 2M}\frac{1}{(\det(I+\rho\theta^2 XX^*))^{n_r}}.
$$

This discussion leads us to consider sums of the type
\begin{equation}\label{siftedsum}
 \sum_{X\in L,\; 0 < ||X||_F\leq M} \frac{1}{(\det(I+c XX^*))^{m}},
\end{equation}
where $c$ is considered a variable.

\begin{remark}
We remark that when 
%the term 
$c$ is very large, 
%(in suitable sense) 
the terms in \eqref{siftedsum} are well-approximated  by $1/\det(c XX^*)^{m}$. In the case $T=n$, we can consider sums of the type
\begin{equation} \label{approximate_sum}
S_L^m(M):=\sum_{X\in L(M)} \frac{1}{|\det(X)|^{m}}.
\end{equation}
The asymptotic behavior of these sums, and its relation to the diversity-multiplexing trade-off of space-time codes, were analyzed in \cite{VLL2013}.
\end{remark}

In this paper, we will address some additional aspects of MIMO space-time code optimization that are not captured by the approximate sums (\ref{approximate_sum}), but instead require to study the original sums (\ref{siftedsum}). In particular, we will consider the following problems:
\begin{enumerate}
\item[-] Find upperbounds of the type $$\sum_{X\in L,\; 0<||X||_F \leq M} \frac{1}{(\det(I+c XX^*))^{m}}\leq c^{-k}f(M)$$ for some function $f$ and positive constant $k$.
\item[-] How large should $m$ be for the  sum  \eqref{siftedsum} to converge?
\item[-] What is the highest power  $k$ such that $$\sum_{X\in L, ||X||_F \leq M} \frac{1}{(\det(I+c XX^*))^{m}}\leq c^{-k} G$$ for some constant $G$ and 
for every $M$?
\end{enumerate}

In the following we will give some general answers to these questions and build a framework for using these sums to analyze codes.

%\section{Some methods for analyzing the shifted determinant sum}
\section{Dyadic summing and upper bounds for shifted inverse determinant sums}
%Let us suppose we have a $v$-dimensional lattice $L$ in  $M_n(\C)$. Let $f$ be a positive valued function from $M_n(\C)$ to  $\R$.

%In the following we will use the notation
%$$
%L(M)=\{x | x\in L, ||x||_F\leq M\}.
%$$
Let's start by considering the decomposition of the shifted determinant. 
Let $\lambda_1,\lambda_2,\ldots,\lambda_n$ be the eigenvalues of $XX^*$. Then 
\begin{eqnarray*}
\det(I+cXX^*)=(1+\lambda_1 c)(1+\lambda_2c)\cdots (1+\lambda_nc)=\\
=1 + \binom{n}{1}p_1 c +\binom{n}{2}p_2c^2 +\cdots + p_nc^n,
\end{eqnarray*}
where $\binom{n}{i}p_i$ is the i-th symmetric polynomial of variables $\lambda_1, \dots, \lambda_n$. One should note that
$$
p_1=\tr(XX^*)=||X||_F^2\quad \mathrm{and}\quad p_n=\det(XX^*).
$$

The following inequalities will be useful in the sequel:
\begin{proposition}[McLaurin's and Newton's inequalities]
The coefficients $p_i$ satisfy
\begin{align}
&p_1\geq \sqrt{p_2}\geq \sqrt[3]{p_3}\geq \cdots \geq \sqrt[n]{p_n}, \label{McLaurin}\\
&p_i^2\geq p_{i-1}p_{i+1}\notag.
\end{align}
\end{proposition}

\begin{corollary}\label{simple}
Let us suppose that $\det(XX^*)\geq 1$.
With the previous notation we have that
$$
p_{k}\geq \sqrt[2^{k-1}]{p_1},
$$
for all $n-1\geq k$.
\end{corollary}
\begin{IEEEproof}
We have that
$$
p_i\geq \sqrt{p_{i+1}p_{i-1}}.
$$
Due to the condition $p_n \geq 1$ we have that $p_k \geq 1 \;\forall k$. Therefore
$$
p_i\geq \sqrt{p_{i-1}}.
$$
Induction now gives us the result.
\end{IEEEproof}

In the following we are interested in asymptotics and convergence and therefore we can forget the binomial terms and concentrate on
the terms $p_i$.  The following inequalities  formalize this approach:
\begin{align*}
&(\det(I+cXX^*))^m=\\
&=\left(1+\binom{n}{1}p_1c +\binom{n}{2}p_2c^2 +\cdots + p_nc^n\right)^m \geq \\
&\geq (1+p_1c +p_2c^2 +\cdots + p_nc^n)^m \geq \\
& \geq (c||X||_F^2+ c^n|\det(XX^*)|)^m= \\
&=\sum_{i=0}^m \binom{m}{i} c^{i+n(m-i)} \norm{X}_F^{2i} \abs{\det(XX^*)}^{m-i}  
\end{align*}
In particular we have 
\begin{align}
&\sum_{X\in L(M)}\frac{1}{(\det(I+c XX^*))^m} \leq  \notag \\
& \leq \sum_{X\in L(M)}\frac{1}{c^{i+n(m-i)} \norm{X}_F^{2i} \abs{\det(XX^*)}^{m-i}} \label{mixed_terms}
\end{align}
for every $0 \leq i \leq m$. 

The following two Lemmas are useful to provide bounds for the sum in equation (\ref{mixed_terms}). 

\begin{lemma}[Dyadic Summing]\label{dyadic}
Let  $f:\R \mapsto \R$ be a positive valued function, and $I \subset \R$ be a discrete set. Suppose that there exist positive constants $K$ and $s$ such that $\forall M \geq 1$,
$$\sum_{x \in I,\; 1 \leq x \leq M} f(x)\leq KM^s.$$ 
We then have that
\begin{align*}
& \sum_{x\in I,\; 1 \leq x\leq M}\frac{f(x)}{x^t} < K_1 & \text{if} \quad t>s,\\
& \sum_{x\in I,\; 1 \leq x\leq M}\frac{f(x)}{x^t} < K_2 \log(M) & \text{if} \quad t=s,\\
& \sum_{x\in I,\; 1 \leq x\leq M}\frac{f(x)}{x^t} < K_3 M^{s-t} & \text{if} \quad t<s,
\end{align*}
%$$
%\sum_{x\in I,\; 1 \leq x\leq M}\frac{f(x)}{x^t}
%$$
%converges if $t>s$. If $t \leq s$ we have an upper bound 
%$$
%\sum_{x \in I,\; 1 \leq x\leq M}\frac{f(x)}{x^t}\leq K_1M^{s-t}\log(M),
%$$
for some constants $K_1$, $K_2$, $K_3$ (depending on $s$ and $t$).
\end{lemma}
\begin{IEEEproof}
By partitioning the interval $[1,M]$ into subintervals of the form $[2^{i-1},  2^i]$, we get
%We begin by dividing the  interval $[1,M]$ into subintervals $
%\{[1, 2^1],\dots (2^{\log_2(M) -1}, 2^{\log_2 (M)]}]\}.$
%Now we divide the whole we sum to subsums where we sum  over the elements in $x\in [2^{i-1},  2^i]$ and then estimate
\begin{align}
&\sum_{x\in I,\; 1 \leq  x \leq M}\frac{f(x)}{x^t}  \leq \sum_{i=1}^{\ceil{\log_2(M)}}\sum_{x\in I,\; 2^{i-1}\leq  x \leq 2^i}\frac{f(x)}{x^t}\leq \notag\\
& \leq  \sum_{i=1}^{\ceil{\log_2(M)}} \sum_{x\in I,\;  2^{i-1} \leq  x \leq 2^i}\frac{f(x)}{2^{(i-1)t}} \leq  \sum_{i=1}^{\ceil{\log_2(M)}}  \frac{K{2^{is}}}{2^{(i-1)t}}=\notag \\  
%Thus, we have shown that
%$$
%\sum_{x\in I,\; 1 \leq  x \leq M}\frac{f(x)}{x^t} \leq 
&=2^t K\sum_{i=1}^{\ceil{\log_2(M)}} (2^{(s-t)i}). \tag*{\IEEEQED}
\end{align}
%which gives the wanted results.
\let\IEEEQED\relax
\end{IEEEproof}

\begin{lemma}\label{reduction}
Let $L\subset M_{n\times T}(\C)$ be a lattice such that $\norm{X}_F \geq 1$ for all the non-zero points $X \in L$. 
Let $g$ be a positive valued function defined in all the non-zero points of the lattice.
If 
$$
\sum_{X\in L(M)} g(X)\leq K M^s
$$
for some fixed  positive constants $K$ and  $s$, then
\begin{align*}
& \sum_{X\in L(M)} \frac{g(X)}{||X||_F^t} < K_1 & \text{if} \quad t>s,\\
& \sum_{X\in L(M)} \frac{g(X)}{||X||_F^t} < K_2 \log(M) & \text{if} \quad t=s,\\
& \sum_{X\in L(M)} \frac{g(X)}{||X||_F^t} < K_3 M^{s-t} & \text{if} \quad t<s,
\end{align*}
for some constants $K_1$, $K_2$, $K_3$.
%
%$$
%\sum_{X\in L(M)} \frac{g(X)}{||X||_F^t}
%$$
%converges if $t>s$. If $t \leq s$ we have
%$$
%\sum_{X\in L(M)} \frac{g(X)}{||X||_F^t}\leq K_1 M^{s-t} \log(M),
%$$
%where $K_1$ is some constant.
\end{lemma}
\begin{IEEEproof}
This is simply the previous proposition applied
%. Here the interval  $[1,M]$ consist of   the radius. The function $f$ is the sum 
to the function $f(x)=\sum_{X\in L,\; ||X||_F=x} g(X)$.
\end{IEEEproof}
Note that the hypothesis $\norm{X}_F \geq 1$ in Lemma \ref{reduction} doesn't incur any loss of generality since we can just rescale the lattice. However, in that case the constants $K_1, K_2, K_3$ will depend on the scaling factor.

%\section{Some basic results on shifted  determinant sums }
%Let us now suppose that we have a lattice $L$  and we have chosen a finite code

%The following result is a consequence of Lemma \ref{reduction}:
%\begin{lemma}\label{killing}
%Let us suppose that  
%$$\sum_{X\in L(M)} \frac{1}{|\det(X)|^{2m}}\leq KM^s,$$
%for some positive constants $s$  and $K$.  We then  have that
%\begin{align*}
%%\sum_{X\in L(M)} \frac{1}{\det(cXX^*)^m(c||X||_F)^t} \leq \frac{G_1}{c^{t+nm}}M^{|s-t|} \quad \mathrm{if} \; s\neq t,
%&\sum_{X\in L(M)} \frac{1}{\det(XX^*)^m ||X||_F^t} \leq G_1 \quad \mathrm{if} \; t > s,\\
%%\sum_{X\in L(M)} \frac{1}{\det(cXX^*)^m(c||X||_F)^t} \leq \frac{G_2}{c^{t+nm}} \log(M) \quad \mathrm{if} \; s= t,
%&\sum_{X\in L(M)} \frac{1}{\det(XX^*)^m||X||_F^t} \leq G_2 \log(M)  \quad \mathrm{if} \; t = s,\\
%&\sum_{X\in L(M)} \frac{1}{\det(XX^*)^m||X||_F^t} \leq G_3 M^{s-t}   \quad \mathrm{if} \; t < s,
%\end{align*}
%for some constants $G_1$, $G_2$ and $G_3$.
%\end{lemma}

We can now obtain a set of upper bounds for shifted inverse determinant sums:
\begin{proposition} \label{Prop3-2}
Let us suppose that $L$ is a $k$-dimensional 
%NVD 
lattice in $M_{n\times T}(\mathbb{C})$, such that $\norm{X}_F \geq 1$ for all the non-zero points $X \in L$, and that we have a bound
$$\sum_{X\in L(M)} \frac{1}{|\det(XX^*)|^{l}}\leq KM^{s(l)}.$$
%$S_L^m(M)\leq KM^{s(m)}$. 
%not the same notation as in the previous proposition!!
We then have that
\begin{eqnarray*}
 \sum_{X\in L(M)}\frac{1}{(\det(I+c XX^*))^m} \leq \min_{0 \leq i \leq m}\{W_i(M)\},
 \end{eqnarray*}
where  for $i\in \{0,\dots, m-1\}$ we have
\begin{align*}
&W_i(M) =  \frac{G_i}{c^{i+n(m-i)}},   &\mathrm{if}\,\, s(m-i)<2i,\\
&W_i(M) = \frac{G_i}{c^{i+n(m-i)}}\log(M), \,\, &\mathrm{if}\,\, s(m-i)=2i,\\
&W_i(M) = \frac{G_i}{c^{i+n(m-i)}} M^{s(m-i)-2i}  &\mathrm{if}\,\, s(m-i)> 2i,
\end{align*}
where $G_i$ are some constants. When $i=m$, we have $W_m(M) = G_m c^{-m}$ if $k < 2m$, $W_m(M) = G_m c^{-m} \log M$ if $k = 2m$ and 
 $W_m(M) = G_m c^{-m} M^{k-2m}M$ if $k > 2m$. 
\end{proposition}

\begin{IEEEproof}
The conclusion follows from equation (\ref{mixed_terms}) and from Lemma \ref{reduction} with $g(X)=1/\det(XX^*)^{m-i}$. For the special case $i=m$, observe that the number of lattice points in $L(M)$ is proportional to the volume of the ball of radius $M$ in $\R^k$. 
\end{IEEEproof}

As a consequence of Proposition \ref{Prop3-2} in the case $i=m$, the shifted determinant sum will converge for $m>k/2$: 

\begin{proposition}[Convergence]
Let us suppose that $L$ is a $k$-dimensional lattice in $M_{n\times T}(\C)$ such that $\norm{X}_F \geq 1$ for all the non-zero points $X \in L$. We then have that
$$
\sum_{X\in L(M)}\frac{1}{(\det(I+c XX^*))^{k/2+\epsilon}}\leq G_{\epsilon}c^{-k/2},
$$
where $\epsilon$ is any positive number and $G_{\epsilon}$ a constant independent of $M$, but dependent on $\epsilon$.
\end{proposition}
%\begin{IEEEproof}
%Let us first estimate
%$$
%\sum_{x\in L(M)}\frac{1}{(\det(I+c XX^*))^{k/2+\epsilon}}\leq \sum_{x\in L(M)}\frac{1}{||X||_F^{k+2\epsilon}}.
%$$
%As we know that $\sum_{x\in L(M)} 1\leq KM^{k}$, for some constant $K$, we can now apply Lemma \ref{reduction} to get the final result.
%\end{IEEEproof}

We can conclude that while 
$$
\sum_{X\in L(M)}\frac{1}{\det(XX^*)^m}
$$
does not usually converge for any $m$ \cite{VLL2013}, quite the opposite is true for the sum
$$
\sum_{X\in L(M)}\frac{1}{(\det(I+c XX^*))^m}.
$$ 
As long as the power $m$ is large enough this sum will always converge.

\section{Shifted inverse determinant sums and SNR level analysis }
Let us introduce one more use for shifted inverse determinant sums. 
%we can derive some conclusions on the performance of space-time codes
If we have a  finite space-time NVD lattice code in $M_n(\C)$,
% where all the differences of codewords are invertible
then in the high SNR regime the diversity order is $nn_r=n_t n_r$. However this regime is rarely visible on error performance curves.  
%Instead we usually see a curve which has quite different slope.  
%We now give some preliminary study, 
We will now see how shifted inverse determinant sums can explain this behavior,
%. We also show how we can 
%measure the point when 
and provide an estimate of the SNR threshold beyond which higher diversity kicks in. 

%The following Proposition shows how the sifted determinant sum bounds provide a new bound for the error performance. 
%With the  previous assumption we have.
\begin{proposition}\label{error}
Suppose that we have a $k$-dimensional lattice $L\in M_{n\times T}(\C)$ such that a determinant sum upper bound 
$$
\sum_{X\in L(M)}\frac{1}{(\det(I+c XX^*))^m} \leq KM^t c^{-d},
$$
holds for some constants $d$,$K$ and $t$. 
%Let us suppose we have a code $L(M)$ in $n\times m$ channel. 
Then the average error probability is upper bounded as
\begin{equation}\label{economical}
P_e( \rho ) \leq  KM^{d+t} \rho^{-d},
\end{equation}
when transmitting with signal-to-noise ratio $\rho$.
\end{proposition}
\begin{IEEEproof}
The average energy of the code $\frac{1}{\sqrt{M}}L(M)$ is less than $1$. Let us now suppose that
$\theta>1 $ is such a constant that $\theta\frac{1}{\sqrt{M}}L(M)$ has average energy 1.
We then have that
\begin{eqnarray}
P_e( \rho ) \leq \sum_{X\in L(M)}\frac{1}{(\det(I+\rho\theta^2 XX^*))^m}\notag \\
\leq KM^{t+d} (\rho\theta^2)^{-d}=  K M^{t+d} {\rho}^{-d}, \label{diversitypoint}
\end{eqnarray}
which concludes the proof.
\end{IEEEproof}

This result has several implications.
%can now be used to several purposes. 
The first is that we can estimate the SNR threshold beyond which we can see diversity order $d$. We can see from equation \eqref{diversitypoint} that when $SNR=\rho\geq K'M^{(t+d)/d}$ the diversity $d$ will appear; before this point we don't have guaranteed diversity $d$. 

Another implication is easier to explain through an example.
\begin{exam}
Let us suppose that we have an $8$-dimensional lattice code $L$ in $M_2(\C)$ and  bounds
$$
\sum_{X\in L(M)}\frac{1}{(\det(I+c XX^*))^4} \leq \min\{ K_1M^4 c^{-8}, K_2c^{-4}\}.
$$
In order to have guaranteed diversity $8$ we must have $\rho\geq K_1'M^{3/2}$. For
guaranteed diversity  $4$ the SNR condition is $\rho\geq K_2'M^{1}$. This shows that when the code grows we need eventually considerably more energy to have guaranteed diversity $8$, independently of the size of the constants $K_1$ and $K_2$. 
\end{exam}

\section{Shifted inverse  determinant sums and DMT analysis}
It was proven in \cite{VLL2013} that the growth of the inverse determinant sums of a lattice code $L\in M_n(\C)$ describes the diversity-multiplexing gain trade-off (DMT) \cite{ZT} of the code $L$ for multiplexing gains $r\in[0,1]$.  We will now show that the shifted determinant sum bounds are useful to analyze the DMT of a code for higher multiplexing gains. We will also show how we can use these sums to evaluate the DMT of a lattice code under naive lattice decoding.

\subsection{Lower bounds for the DMT under ML decoding}

\begin{definition}
Given the lattice $L \subset M_{n\times T}(\C)$, a space-time lattice coding scheme associated with $L$ is a collection of STBCs where each member is given by
\begin{equation}\label{codingscheme}
C_L( \rho)=\rho^{-\frac{rT}{k}}L\left(\rho^{\frac{rT}{k}}\right)
\end{equation}
for the desired multiplexing gain $r$ and for each $\rho$ level.
\end{definition}

\begin{proposition}
Let $L$ be a $k$-dimensional lattice in $M_{n\times T}(\C)$  and suppose that the determinant sum upper bound 
$$
\sum_{X\in L(M)}\frac{1}{(\det(I+c XX^*))^m} \leq K c^{-a} f(M),
$$
holds for some positive constants $K$ and $a$.  We then have that for SNR $\rho$  the average error probability of the code $C_L(\rho)$
%$L(\rho^{\frac{rT}{k}})$ 
has an upperbound 
$$
P_e(\rho)\leq K_1 \rho^{-a+2arT/k}f(2\rho^{\frac{rT}{k}}).
$$
\end{proposition}
\begin{IEEEproof}
The average energy for the code $C_L(\rho)$is less than $1$. Now for transmission with SNR
$\rho$, each of the codewords gets multiplied with $\rho^{1/2}$ as in (\ref{eq:channel}). We now have
$$
P_e(\rho) \leq \sum_{X\in L(2\rho^{\frac{rT}{k}})}\frac{1}{(\det(I+c XX^*))^m},
$$
where $c=\rho^{1-2rT/k}$. The final result is then simply gotten by substitution.
\end{IEEEproof}

The following DMT bound is a direct corollary of the previous result:
\begin{corollary} \label{DMT_corollary}
Let us suppose  that we have an upperbound
$$
\sum_{X\in L(M)}\frac{1}{(\det(I+c XX^*))^m} \leq K c^{-a}M^b.
$$
We then have that the DMT of the code $L$ is lowerbounded by the following line:
$$
[r,(a-rT(2a+b)/k )^+ ].
$$
\end{corollary}

We also have the following curiosity, which shows that any full dimensional lattice achieves the full multiplexing gain, when we have enough receive antennas:

\begin{corollary} 
Let $L$ be a $2nT$-dimensional lattice code in $M_{n\times T}(\C)$. If $n_r>nT+1$, 
the  code 
%$L(\frac{rT}{k})$ 
$C_L(\rho)$
has an upperbound for the average error probability
$$
P_e\leq \rho^{(1-r/n)n_r},
$$ 
when transmitting with SNR $\rho$.
\end{corollary}
\begin{IEEEproof}
This follows from Proposition \ref{Prop3-2} with $m=n_r>k/2$, yielding $a=n_r$ and $b=0$ in Corollary \ref{DMT_corollary}. 
\end{IEEEproof}

\subsection{Lower bound for the DMT under naive lattice decoding}
Let us consider naive lattice decoding as defined in \cite{TK}, which consists in minimizing the Euclidean metric with respect to the received signal over all the lattice points $X' \in L$, regardless of whether they belong to the finite code. 
%We will again use the same normalization as in the previous section, but we  will multiply the whole lattice with normalization term that corresponds to the code $L(\rho^{\frac{rT}{k}})$.  We do not claim that our approach provides optimal bounds;
%given is particularly good tool for analyzing this setting. 
%rather this is an invitation to develop our suggestion further. We begin with a small introduction.

%Let us first forget the questions of energy normalization. 
%Naive lattice decoding, when transmitting a codeword $X_i \in L\subseteq M_{n\times T}(\C)$, 
%we consider minimization of
%consists in minimizing
%$$
%||HX_i+N-HX_j||_F, 
%$$
%where $N$ is the noise, over all the lattice points $X_j \in L$, $i\neq j$.  In this context a decoding error will occur 
%we will make a mistake in receiving 
%if the received point is closer to  some $HX_j$ in $HL$ than $HX_i$. The usual determinant criterion gives an upper bound for the  average probability  that the received point is closer to a fixed $X_j$ than intended $X_i$. 
It is clear that the average probability of error of naive lattice decoding can be upper bounded by a determinant sum over the whole lattice. This bound is relevant only when the sum is converging. We thus state the following Proposition, which can be proven in the same way as Corollary \ref{DMT_corollary}:

\begin{proposition}\label{naive}
Let $L\in M_{n\times T}(\C)$ be a $k$-dimensional lattice, and suppose that a determinant sum upper bound 
\begin{equation}\label{infinitebound}
\sum_{X\in L(M)}\frac{1}{(\det(I+c XX^*))^m} \leq K c^{-a},
\end{equation}
holds for some positive constants $K$ and $a$.  Then the DMT of the code $C_L(\rho)$, under naive lattice decoding,  is lower bounded by
$$
[r,(a-2rTa/k)^+ ].
$$
\end{proposition}

%\begin{IEEEproof}
%Just  as before the finite code is  $\rho^{1/2-\frac{rT}{k}}L(\rho^{\frac{rT}{k}})$, but in order to perform naive lattice decoding we multiply the whole lattice $L$ with $\rho^{1/2-\frac{rT}{k}}$. 
%We can upperbound the error probability under naive lattice decoding by using the union bound over the whole lattice. We then have that
%$$
%p_e(\rho) \leq \sum_{x\in L }\frac{1}{(\det(I+c XX^*))^m},
%$$
%where $c=\rho^{1-rT2/k}$. The final result is then simply gotten by substitution.
%
%\end{IEEEproof}

\section{Examples}

\subsection{Analyzing the Golden code}
Let us consider the Golden code $L \subset M_2(\C)$, which is an $8$-dimensional lattice.
According to \cite{ITW2013}, if $n_r>1$ we have
\begin{equation}\label{growth}
\sum_{X\in L(M)}  \frac{1}{|\det(X)|^{2n_r}}\leq KM^4,
\end{equation}
where $K$ is a positive constant.

%\begin{proposition}
With the previous notation we have that
$$
\sum_{X\in L(M)}\frac{1}{(\det(I+c XX^*))^4}\leq \min_{i \in \{0,2,4\}}W_i(M),
$$
where $W_0 = K_0 c^{-8}M^4$, $W_2= K_2 c^{-6}\log(M)$ and  $W_4= K_4 c^{-4}$ for some constants $K_i$. 
%\end{proposition}

%We can now see that the economical diversity order for Golden code, when received with $4$ antennas, is at least $4$. 
%However, the upper bound $W_2$ and diversity order $6$ is very close.

\begin{proposition}
Under naive lattice decoding, when received with $4$ antennas, the Golden code achieves the DMT curve
$$
[r, (2(2-r))^+ ].
$$
With ML the Golden code achieves the DMT curve
$$
[r, \mathrm{max}\{(8-5r),(6-3r)\}^+],
$$
which coincides with the optimal DMT $[r, (4-r)(2-r)^+]$.
\end{proposition}

We can see that even with naive lattice decoding the Golden code does achieve the optimal multiplexing gain. However, the maximal diversity is only
$4$.

\begin{remark}
 Here one should note that for multiplexing gains $r\in [0,1]$ the sum 
$$
\sum_{X\in L(M)}\frac{1}{(\det(c XX^*))^4},
$$
does provide the best upper bound. However,  when
 $r\in [1,2]$ the sum
$$
\sum_{X\in L(M)}\frac{1}{(c^6||X||_F^4 \mathrm{det}(XX^*)^2)},
$$
gives a tighter upper bound.

\end{remark}

\subsection{Analyzing diagonal number field codes}

Let us now consider a complex diagonal number field code. Such a code is $2n$-dimensional NVD lattice in $M_n(\C)$. As proved in \cite{VLL2013} we have that for $m\geq 1$ we have
\begin{equation}\label{diagonal_bound}
\sum_{X\in L(M)}\frac{1}{|(\det(XX^*)|^m}\leq K \log(M)^{3n-1},
\end{equation}
for some constant $K$.

\begin{proposition}
Let $L$ be a diagonal number field code in $M_n(\mathbb{C})$ such that $\det(XX^*) \geq 1$ for all $X \in L,\; X \neq \mathbf{0}$ and let $m>1$. We then have that
$$
\sum_{X\in L(M)}\frac{1}{(\det(I+c XX^*))^m} \leq Kc^{-nm+1},
$$
where $K$ is some constant independent of $M$.
\end{proposition}
\begin{IEEEproof}
We begin with
$$
\sum_{X\in L(M)}\frac{1}{(\det(I+c XX^*))^m} \leq \sum_{X\in L(M)}\frac{1}{(c^np_n+c^{n-1}p_{n-1})^m}
$$
$$
\leq c^{-mn+1}\sum_{X\in L(M)}\frac{1}{(p_n^{m-1}p_{n-1})}
$$
$$
\leq c^{-mn+1}\sum_{X\in L(M)}\frac{1}{(p_n^{m-1}(p_1^{1/{2^{(n-2)}}}))},
$$
where the last equation follows from Corollary \ref{simple}. As $p_1=||X||_F^2$ we can then apply  Lemma \ref{reduction}. 
%\ref{killing}.
\end{IEEEproof}

As a Corollary to the previous we have the following:

\begin{proposition}
Under naive lattice decoding  the number field code achieves the DMT curve
$$
[r, ((n_tn_r-1)(1-r))^+].
$$
\end{proposition}

\section*{Acknowledgement}
 The authors are grateful for Eeva Suvitie for suggesting the use of dyadic summing.
The research of  R. Vehkalahti is supported  by the Academy of
Finland  grant  \#252457.

\end{document}